\documentclass[aps,prl,superscriptaddress, twocolumn,nolongbibliography]{revtex4}

\usepackage{amsmath}
\usepackage{amssymb}
\usepackage{color}
\usepackage{enumitem}
\usepackage{graphicx}
\usepackage{epsfig}

\newcommand{\euler}{\mathrm{e}}
\newcommand{\iu}{\mathrm{i}}

\begin{document}
	
	\newcommand{\jav}[1]{\textcolor{red}{#1}}
	
	\title{Unified Description of the Aharonov--Bohm Effect in Isotropic Multiband Electronic Systems}
	
	\author{R\'obert N\'emeth }
	
	\author{ J\'ozsef Cserti}
	\email[Corresponding author e-mail:]{cserti@elte.hu}
	
	\affiliation{Department of Physics of Complex Systems,
		ELTE E{\"o}tv{\"o}s Lor{\'a}nd University
		\\ H-1117 Budapest, P\'azm\'any P{\'e}ter s{\'e}t\'any 1/A, Hungary}
	
	\begin{abstract}
		
		We present a unified treatment of the Aharonov--Bohm (AB) effect for two-dimensional multiband electronic systems possessing isotropic band structures. We propose an integral representation of the AB scattering state of an electron scattered by an infinitely thin solenoid. Moreover, we derive the asymptotic form of the AB scattering state and obtain the differential cross section from that. We found a remarkable result, namely that  
		this cross section is {\it the same for all isotropic systems} and agrees with that obtained first by Aharonov and Bohm for spinless free particle systems.
		To demonstrate the generality of our theory, we consider several specific multiband systems relevant to condensed matter physics. 
		
	\end{abstract}
	
	%\pacs{71.70.Ej, 73.63.Hs, 81.05.Uw, 73.43.Cd}
	
	\maketitle
	
	Aharonov and Bohm (AB) in their seminal paper~\cite{Aharonov-Bohm:cikk} calculated how the incident plane wave of a spinless free particle is scattered by an infinitely thin magnetic solenoid and the differential scattering cross section. 
	The first positive observations of this quantum effect were reported by Chambers~\cite{PhysRevLett.5.3}, Tonomura~\emph{et al.}~\cite{Tonomura_PhysRevLett.48.1443},  
	Web~\emph{et al.}~\cite{Webb_PhysRevLett.54.2696} and Ensslin and coworkers~\cite{Fuhrer_Ensslin_2001,Ensslin_PhysRevLett.93.066802,ENSSLIN2005585}. 
	For a review, see Olariu and Popescu's work~\cite{Popescu_RevModPhys.57.339}. 
	Further theoretical work~\cite{Hagen_PhysRevD.41.2015} %~\cite{RUIJSENAARS19831,Hagen_PhysRevD.41.2015} 
	clarified the subtle issue regarding the scattering amplitude in the forward direction. 
	The concept of whirling states introduced by Berry~\cite{Berry_1980} has been proved to be an alternative procedure for constructing the AB scattering state. 
	%The reality of the Aharonov--Bohm effect is now firmly established, and several textbooks have already mentioned \jav{Én itt is inkább jelen időt használnék.} the effect (see, e.g., Ref.~\cite{Sakurai:1341875} \jav{Kell a ``Ref.''? Máshol nincs.}).
	
	The scattering of a relativistic fermion off a vortex in 2+1 dimensions as an extension of the original work~\cite{Aharonov-Bohm:cikk} has been studied first by Alford and Wilczek~\cite{Wilczek_PhysRevLett.62.1071} and subsequently discussed in Refs.~\cite{Gerbert_PhysRevD.40.1346,Hagen_PhysRevLett.64.503,Hagen_PhysRevD.42.3524,PhysRevB.82.075316,PhysRevB.84.125306}.
	Furthermore, the whirling state idea introduced by Berry~\cite{Berry_1980} has been generalized to the relativistic regime by Girotti and Romero~\cite{Girotti_1997}.
	More recently, the Aharonov--Bohm  interferences in a usual two-slit-like setup have been studied in single layer graphene~\cite{Carlo_AB_Graphene_PhysRevB.76.235404,AB_Graphene_PhysRevB.77.085413,AB_Graphene_Huefner_2009,AB_Graphene_Huefner_2010,AB_graphene_Wurm_Richter_2010,AB_graphene_Vozmediano_deJuan_2011}  and in bilayer graphene~\cite{PARK20171831}. The conventional AB scattering problem has also been studied recently in graphene~\cite{Ramezani_Peeters_scatt_2013,Rycerz_PRB.101.245429}.
	Magnetic scattering of Dirac fermions in topological insulators and graphene was investigated in Ref.~\cite{Dirac_in_top_insul_and_graphene_Egger_PhysRevB.82.155431}. 
	
	In the present work, we extend the Aharonov--Bohm scattering problem to a broader class of  Hamiltonians. In particular, for isotropic multiband Hamiltonians, we present an integral representation of the AB scattering state in which the electron is scattered by an idealized, infinitely thin solenoid carrying a flux $\Phi$.
	We shall show rigorously that our proposed AB scattering state constructed from the energy eigenstates in the absence of a magnetic field satisfies the Schr\"odinger equation of the electron scattering off a flux line.
	Moreover, from the asymptotic form of the AB scattering state, we found a remarkable result for the differential cross section, namely, it is the same for all isotropic multiband systems. 
	Note that as a special case of our work, Alford and Wilczek found the same results  for Dirac electrons~\cite{Wilczek_PhysRevLett.62.1071}. 
	
	The most general Hilbert space of the systems we study in this work takes 
	the form $\mathcal{H} = L^2(\mathbb{R}^2,\mathbb{C})\otimes\mathbb{C}^D$ consisting of the two-dimensional spatial and a $D$-dimensional internal degrees of freedom, for instance, spin or isospin. Furthermore, we limit ourselves to studying Hamiltonians satisfying the following requirements:
	
	\begin{itemize}
		
		\item\emph{Polynomicity}. The Hamiltonian of the system is given as a polynomial of the momentum operators $\hat{p}_x$ and $\hat{p}_y$ with degree $(I+J)$:
		\begin{equation}
			\hat{H} = \sum_{i = 0}^I \sum_{j = 0}^J \hat{p}_x^{i} \hat{p}_y^{j} \otimes \hat{T}_{ij} ,
			\label{H-matrix:eq}
		\end{equation} where $\hat{T}_{ij}$ is a  $D \times D$ hermitian matrix for each $i,j$. 
		A few well-known examples of such Hamiltonians are listed in Table~\ref{systems:table}.
		
		\item\emph{Isotropy}. In the absence of a magnetic field the eigenfunction of the Hamiltonian (\ref{H-matrix:eq}) as a plane wave
		solution of the Schr\"odinger equation with energy $E_{s}(\boldsymbol{k})$  propagating in the direction $\boldsymbol{k}$  takes the form
		\begin{equation}
			\boldsymbol{\Psi}_{s,\boldsymbol{k}}(\boldsymbol{r}) = \euler^{\iu \boldsymbol{k r}} \, \mathbf{u}_{s}(\boldsymbol{k}) 
			= \euler^{\iu kr\cos(\varphi - \vartheta)}\, \mathbf{u}_s(k,\vartheta), 
			\label{plane_wave_no_B:eq}
		\end{equation}
		where 
		$\boldsymbol{k} = k {\left[\cos \vartheta, \sin \vartheta\right]}$ is the wavenumber in polar coordinates  $(k,\vartheta)$,
		$\boldsymbol{r} = r {\left[\cos \varphi, \sin \varphi\right]}$ is the position in polar coordinates $(r,\varphi)$,  $\mathbf{u}_s(\boldsymbol{k})$ is a $D$-component vector and $s =1,\dots, D$ labels the energy band.
		Here we assume that the  dispersion relation $E_{s}(\boldsymbol{k})$ for all bands $s$ is isotropic, i.e., it depends only on the magnitude $|\boldsymbol{k}|$.
		
		\item\emph{Regularity}. At the flux line, the regularity of at least one component of the AB scattering state is required.  This is a purely mathematical assumption which, however, has been physically justified in several special cases~\cite{Gerbert_Jackiw_MIT_Report_1989,Hagen_PhysRevD.42.3524,Hagen_PhysRevLett.64.503,
			Gerbert_PhysRevD.40.1346,Wilczek_PhysRevLett.62.1071}. Relaxation of this requirement might also become possible in future generalizations of our method.
		
	\end{itemize} We should emphasize that the above constraints on the multiband systems still allow a very wide class of Hamiltonian operators.
	
	For a flux line with magnetic flux $\Phi$ along the $z$-axis, the vector potential 
	in symmetric gauge reads
	\begin{equation}
		\boldsymbol{A}(x,y) = \frac{\Phi}{2\pi (x^2+y^2)} \begin{pmatrix}
			-y \\ x
		\end{pmatrix} . 
		\label{Dirac_A:def}
	\end{equation}
	Thus, the Hamiltonian \eqref{H-matrix:eq} must be modified such that the momentum operators $\hat{p}_x$ and $\hat{p}_y$ are replaced by 
	$\hat{\Pi}_x = \hat{p}_x +  e A_x(\hat{x},\hat{y})$ and $\hat{\Pi}_y = \hat{p}_y +  e A_y(\hat{x},\hat{y})$ (here $e>0$ is the magnitude of the electron charge): 
	\begin{equation}
		\hat{H} = \sum_{i = 0}^I \sum_{j = 0}^J \hat{\Pi}_x^{i} \hat{\Pi}_y^{j} \otimes \hat{T}_{ij} .
		\label{H-magnetic:eq}
	\end{equation}
	
	To obtain the general solution of an incident electron scattered by the flux tube in multiband systems we construct it from the 
	plane wave solution (\ref{plane_wave_no_B:eq}) propagating along the direction $\vartheta$, in the following way
	\begin{align}
		\boldsymbol{\Psi}^{(+)}_{s,\boldsymbol{k}}(r,\varphi) &= 
		\sum_{m = -\infty}^{\infty} \frac{\epsilon(m+\alpha)}{2\pi}  \nonumber \\
		\times & \int\displaylimits_{\Gamma(m+\alpha,\varphi)}\hspace{-12pt}\mathrm{d}\xi~ \boldsymbol{\Psi}_{s,\boldsymbol{K}}(r,\varphi) \euler^{\iu m(\xi - \vartheta) - \iu\alpha(\varphi - \xi)},
		\label{tot_scatt_alpha:eq}
	\end{align}
	where $\alpha = \Phi/\Phi_0$ and $\Phi_0 = h/e$ is the flux quantum, while $\epsilon(x)$ is the sign function defined as $\epsilon(x) =1$, if $x\ge 0$ and $\epsilon(x) =-1$, if $x< 0$. Furthermore, the wave number in the plane wave $\boldsymbol{\Psi}_{s,\boldsymbol{k}}$ is replaced by 
	a complex wave number as $\boldsymbol{k}\rightarrow \boldsymbol{K} = k {\left[\cos \xi, \sin \xi\right]}$, where $\xi$ is defined in the complex plane.
	The integration contours $\Gamma(m+\alpha,\varphi)$ are curves on the complex plane depending on the sign of $m+\alpha$ and the value of the real space polar angle $\varphi$:
	\begin{equation}
		\Gamma(m+\alpha,\varphi) = \begin{cases} \Gamma_+(\varphi),~\mathrm{if}~m+\alpha\ge 0, \\ \Gamma_-(\varphi),~\mathrm{if}~m+\alpha< 0. \end{cases}
		\label{contours:eq}
	\end{equation}
	The curves $\Gamma_+(\varphi)$ and $\Gamma_-(\varphi)$ further depend on the sign of the radial component $v_{s,k}$ of the group velocity 
	\begin{equation}
		\boldsymbol{v}_{s} = \frac{1}{\hbar} \frac{\partial E_{s}}{\partial \boldsymbol{k}}
		\label{group_velocity:eq}
	\end{equation}
	in $\boldsymbol{k}$-space.
	
\begin{figure}[hbt]
		\centering
		\includegraphics[scale=0.75]{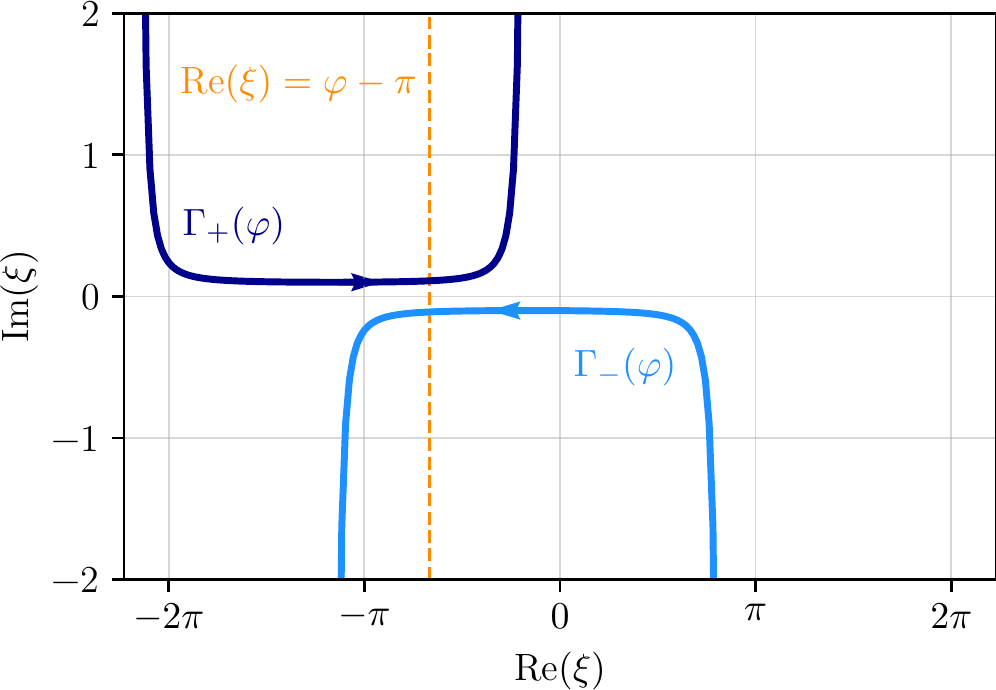}\vspace{10pt}
		\includegraphics[scale=0.75]{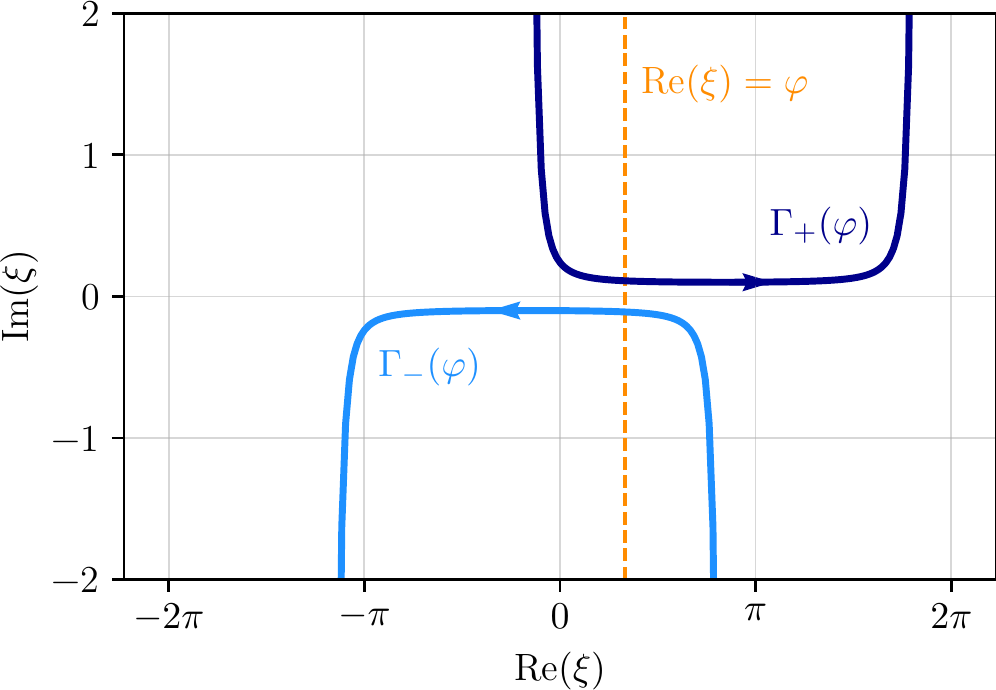}
		\caption{An illustration of the contours $\Gamma_+(\varphi)$ and $\Gamma_-(\varphi)$ in the complex 
			$\xi$-plane  for $v_{s,k}(k)>0$ (top) and $v_{s,k}(k)<0$ (bottom), and $\varphi = \pi/3$. 
			\label{contour1:fig} 
		}
	\end{figure}

	In particular, if $v_{s,k}>0$, the curve $\Gamma_+(\varphi)$ is $\cup$-shaped running from $\xi = -5\pi/2 + \varphi + \iu\infty$ to $\xi = -\pi/2 + \varphi + \iu\infty$ with $\mathrm{Re}(\xi)>0$, and the curve $\Gamma_-(\varphi)$ is $\cap$-shaped running from $\xi = -3\pi/2 + \varphi - \iu\infty$ to $\xi = \pi/2 + \varphi - \iu\infty$ with $\mathrm{Re}(\xi)<0$. However, if $v_{s,k}<0$, the curve $\Gamma_+(\varphi)$ must be shifted by $2\pi$ along the real axis compared to the previous definition. Furthermore, if $v_{s,k}=0$, the AB scattering state has no physical meaning as the corresponding plane waves have constant zero current density indicating that they do not propagate. Such curves are shown in Fig.~\ref{contour1:fig}.
	Note that these `$\cup$-shaped' ($\Gamma_+(\varphi)$) and `$\cap$-shaped'  ($\Gamma_-(\varphi)$) contours stem from one of the integral representations of the Bessel function~\cite{Grads:book}. 
	
	The wave function $\boldsymbol{\Psi}^{(+)}_{s,\boldsymbol{k}}$ proposed in Eq.~(\ref{tot_scatt_alpha:eq}) must satisfy the following conditions:
	\begin{enumerate}[label=(\roman*), leftmargin=1cm]
		\item \label{itm:1} well-defined (convergent and single-valued);
		\item \label{itm:2} eigenvector of the Hamiltonian \eqref{H-magnetic:eq} with eigenvalue $E_s(\boldsymbol{k})$;
		\item \label{itm:3} its asymptotic form (as $r\to\infty$) is the sum of an incoming plane wave  $\boldsymbol{\Psi}_{s,\boldsymbol{k}}^\mathrm{in}$ and an outgoing cylindrical wave $\boldsymbol{\Psi}_{s,\boldsymbol{k}}^\mathrm{out}$.
	\end{enumerate}
	Our detailed proof of these conditions is presented in the Supplemental Material.
	In summary, one of the central results is the exact solution $\boldsymbol{\Psi}^{(+)}_{s,\boldsymbol{k}}$ of the scattering by a flux line given by Eq.~(\ref{tot_scatt_alpha:eq}) 
	constructed from the plane wave solution $\boldsymbol{\Psi}_{s,\boldsymbol{k}}$ in the absence of a magnetic field given by Eq.~\eqref{plane_wave_no_B:eq}. Reversely, for $\alpha=0$ the scattering state is reduced to the plane wave solution.
	
	Similarly to the work by Aharonov and Bohm~\cite{Aharonov-Bohm:cikk}, we now calculate the differential cross section based on condition \ref{itm:3} mentioned above. We obtain the asymptotic form of the wave function given by Eq.~(\ref{tot_scatt_alpha:eq}) using the saddle point approximation~\cite{Berry_Chambers_1980,Arfken}.
	In particular, we find that the incoming wave propagating in direction $\boldsymbol{k} = k \left[\cos \vartheta, \sin \vartheta\right]$ 
	and the outgoing wave are 
	\begin{subequations}
		\label{in_out_wave:eq} %%%%
		\begin{align}
			\boldsymbol{\Psi}_{s,\boldsymbol{k}}^\mathrm{in}(r,\varphi) &= \euler^{\iu kr\cos(\varphi - \vartheta)}\euler^{-\iu\alpha(\varphi - \vartheta + \pi)}\mathbf{u}_s(k,\vartheta), \\ 
			\boldsymbol{\Psi}_{s,\boldsymbol{k}}^\mathrm{out}(r,\varphi) &= \frac{F(\varphi - \vartheta)}{\sqrt{r}}\euler^{\pm\iu kr}\mathbf{u}_s(k,\varphi), 
			\label{asymptotic_outgoing:eq}
		\end{align} %%%
		where the sign $\pm$ corresponds to whether the radial group velocity is positive or negative, respectively, while the \emph{scattering amplitude} $F$ for positive radial group velocity, i.e., $v_{s,k} >0$ is given by 
		%As we shall see, Eq. \eqref{asymptotic_outgoing:eq} ensures that the particle current flows in parallel to the direction $\pm\vartheta$ and flows outwards.  
		%Here the extra phase term $\euler^{-\iu\alpha(\varphi - \vartheta + \pi)}$ in the incoming wave ensures that the current density is parallel to the direction $\vartheta$ for the state $\boldsymbol{\Psi}_{s,\boldsymbol{k}}^\mathrm{in}$. 
		\begin{align}
			F(\varphi - \vartheta) &= \frac{\sin(\alpha\pi)\euler^{-\iu\alpha\pi} \euler^{\iu\left(\lfloor\alpha\rfloor + \frac12\right)(\vartheta - \varphi)}}{\sqrt{2\pi\iu k} \sin\left(\frac{\varphi - \vartheta}{2}\right)}.
			\label{F_ampl:eq} %%
		\end{align} %%%
	\end{subequations}
	Similarly, we calculated the scattering amplitude for negative radial group velocity, i.e., for $v_{s,k} <0$. The detailed derivations of Eq.~(\ref{in_out_wave:eq}) are in the Supplemental Material. 
	
	Finally, for our multiband systems, the differential cross section can be written as
	\begin{equation}
		\sigma_\mathrm{AB}(\varphi) \equiv \frac{\mathrm{d}\sigma}{\mathrm{d}\varphi}(s,\boldsymbol{k};\varphi) = \lim_{r\to\infty} r \frac{|\boldsymbol{j}_{s,\boldsymbol{k}}^{\mathrm{out}}(r,\varphi)|}{|\boldsymbol{j}_{s,\boldsymbol{k}}^\mathrm{in}(r,\varphi)|}, 
		\label{sigma_def:eq}
	\end{equation} where $\boldsymbol{j}_{s,\boldsymbol{k}}^{\mathrm{in}}$ and $\boldsymbol{j}_{s,\boldsymbol{k}}^{\mathrm{out}}$ are the particle current densities corresponding to $\boldsymbol{\Psi}_{s,\boldsymbol{k}}^\mathrm{in}$ and $\boldsymbol{\Psi}_{s,\boldsymbol{k}}^\mathrm{out}$ given by Eq.~\eqref{in_out_wave:eq}, respectively, derived from the Schrödinger equation. Then, after some straightforward algebra detailed in the Supplemental Material, we obtain
	\begin{subequations}
		\label{currents_in_out:eq}
		\begin{align}
			\boldsymbol{j}_{s,\boldsymbol{k}}^\mathrm{in}(r,\varphi) &= \boldsymbol{v}_s(k,\vartheta) , \\
			\boldsymbol{j}_{s,\boldsymbol{k}}^\mathrm{out}(r,\varphi) &= \frac{|F(\varphi - \vartheta)|^2}{r} \boldsymbol{v}_s(k,\varphi) + O\big(r^{-3/2}\big) .
		\end{align}
	\end{subequations} Using Eqs.~(\ref{sigma_def:eq}) and (\ref{currents_in_out:eq}), the $O(r^{-3/2})$ terms vanish in the limit $r\to\infty$, and the differential scattering cross section becomes
	\begin{align}
		\sigma_\mathrm{AB}(\varphi) 
		&= |F(\varphi - \vartheta)|^2 
		= \frac{\sin^2(\alpha\pi)}{2\pi k \sin^2\left[(\varphi - \vartheta)/2\right]}.
		\label{sigma_AB_gen:eq}
	\end{align}
	This is the famous result obtained first by Aharonov and Bohm~\cite{Aharonov-Bohm:cikk} 
	for spinless charged particles (in their case, $\vartheta = \pi$).
	However, we should emphasize that our result is a more general one, valid for \emph{all isotropic systems}, and independent of the band label $s$ as well. 
	
	\begin{widetext}
		\begin{table*}[htb]
			%\begin{tabular}{|c|c|@{\hspace{3mm}} c @{\hspace{3mm}}|
			%@{\hspace{3mm}} c|@{\hspace{3mm}} c @{\hspace{3mm}} |
			%@{\hspace{3mm}} c @{\hspace{3mm}}|} \hline \hline
		%\begin{tabular}{|c|c|p{53mm}|p{35mm}|c|c|} \hline \hline
		\begin{tabular}{lccccl} \hline \hline
			System 
			& $D$ 
			& $\hat{H}$ 
			& $\mathbf{u}_s(k,\vartheta)$
			& $\mathbf{U}_{s,k,m}(r,\varphi)$ for $s=1$
			& Refs.   \\ \hline
			2D electron gas 
			& 1 
			& $ \frac{\hat{p}_x^2 + \hat{p}_y^2}{2M}$   
			& 1 
			& $J_{|m+\alpha|}(k r)\euler^{\iu m \varphi}$ 
			& \cite{datta_1995}
			\\[2ex]
			Monolayer graphene 
			& 2 
			& $  v \left(\hat{p}_x\otimes\hat{\sigma}_x + \hat{p}_y\otimes\hat{\sigma}_y\right)$   
			& $\frac{1}{\sqrt2} \begin{pmatrix} 1 \\ s\, \euler^{\iu \vartheta} \end{pmatrix}$
			& $ \frac{\euler^{\iu m \varphi}}{\sqrt2} \begin{pmatrix} J_{|m+\alpha|}(k r) \\ \iu \epsilon(m + \alpha) J_{|m+\alpha| + \epsilon(m+\alpha)}(k r) \euler^{\iu\varphi} \end{pmatrix}$ 
			& \cite{Novoselov2005,Zhang2005,DiVincenzo_Mele_PRB_29_1685:ref,RevModPhys.81.109}
			\\[4ex]
			Bilayer graphene 
			& 2 
			& $  - \frac{\left(\hat{p}_x^2 - \hat{p}_y^2\right)\otimes\hat{\sigma}_x + 2\hat{p}_x\hat{p}_y\otimes\hat{\sigma}_y}{2M}$   
			& $\frac{1}{\sqrt2} \begin{pmatrix} 1 \\ -s\, \euler^{2 \iu \vartheta} \end{pmatrix}$
			& $\frac{\euler^{\iu m \varphi}}{\sqrt2} \begin{pmatrix} J_{|m+\alpha|}(k r) \\ J_{|m+\alpha| + 2\epsilon(m+\alpha)}(k r) \, \euler^{2\iu\varphi} 
			\end{pmatrix} $ 
			& \cite{Novoselov_Hall:ref,mccann:086805}
			\\[4ex]
			Rashba system
			& 2 
			& $\frac{\hat{p}_x^2 + \hat{p}_y^2}{2M}\otimes \hat{I} 
			+ v \left(\hat{p}_y\otimes\hat{\sigma}_x - \hat{p}_x\otimes\hat{\sigma}_y\right) $
			&$\frac{1}{\sqrt2} \begin{pmatrix} 1 \\ - \iu \, s\, \euler^{\iu \vartheta} \end{pmatrix}$
			& $\frac{\euler^{\iu m \varphi}}{\sqrt2} \begin{pmatrix} J_{|m+\alpha|}(k r) \\ \epsilon(m + \alpha) J_{|m+\alpha| + \epsilon(m+\alpha)}(k r) \euler^{\iu\varphi} \end{pmatrix} $ 
			& \cite{Rashba_cikk_1960,Yu_A_Bychkov_1984,Fabian_RevModPhys.76.323,Winkler:684956}
			\\[4ex]
			Pseudospin-1 system
			& 3
			& $  v \big(\hat{p}_x\otimes \hat{\tau}_x + \hat{p}_y\otimes \hat{\tau}_y\big)$   
			&$\frac{1}{2} \begin{pmatrix} \euler^{-\iu \vartheta} \\ \sqrt{2} \, s\\ 
				\euler^{\iu \vartheta} \end{pmatrix}$
			& $\frac{\euler^{\iu m \varphi}}{ 2\iu} \begin{pmatrix} \epsilon(m + \alpha) J_{|m+\alpha| - \epsilon(m+\alpha)}(k r) \euler^{-\iu\varphi} \\ \sqrt2 \, \iu ~ J_{|m+\alpha|}(k r) \\ \epsilon(m + \alpha) J_{|m+\alpha| + \epsilon(m+\alpha)}(k r) \euler^{\iu\varphi} \end{pmatrix} $ 
			& \cite{PhysRevB.84.115136,PhysRevB.81.041410,PhysRevA.80.063603,PhysRevB.84.195422}
			\\
			\hline \hline
		\end{tabular}
		\caption{The Hamiltonian of a few systems and the partial wave function  $\mathbf{U}_{s,k,m}$ in magnetic field obtained from Eq.~(\ref{tot_scatt_alpha:eq}).
			Here  $D$ is the number of internal degrees of freedom, $\hat{I}$ is a $2 \times 2$ identity matrix, 
			$\hat{\sigma}_x$ and $\hat{\sigma}_y$ are the Pauli matrices,  
			$\hat{\tau}_x$ and $\hat{\tau}_y$ are the corresponding spin-$1$  matrices. 
			More details of these systems can be found in the references listed
			in the last column.
			\label{systems:table}}
	\end{table*}
\end{widetext}	

\textit{Applications: }
Now, using our general integral representation of the AB scattering state given by Eq.~(\ref{tot_scatt_alpha:eq}), we calculate explicitly the wave function for a few well-known systems listed in Table~\ref{systems:table}. 
Actually, using Eq.~(\ref{tot_scatt_alpha:eq}) we find that the AB scattering state can be written in a compact form as
\begin{equation}
	\boldsymbol{\Psi}^{(+)}_{s,\boldsymbol{k}}(r,\varphi) = \sum_{m=-\infty}^\infty \hspace{-5pt}  {\left(-\iu \right)}^{|m+\alpha|} \euler^{\iu m(\pi - \vartheta)}  \mathbf{U}_{s,k,m}(r,\varphi),
	\label{gen_scatstate:eq}
\end{equation}
where the $D$-component partial wave function $\mathbf{U}_{s,k,m}$ for different systems is listed in Table~\ref{systems:table}. 
For details, see the Supplemental Material, where derivations and visualization of the scattering states are presented.
Note that these wave functions are indeed solutions of the Schrödinger equation corresponding to the Hamiltonian \eqref{H-magnetic:eq}. 
Moreover, we should stress that one could apply our approach to other isotropic multiband systems not listed in Table~\ref{systems:table}. Even for experimentally relevant situations such as strained or gapped systems, the scattering cross section does not change provided that the dispersion relation remains isotropic.

In summary, we proposed an integral representation of the scattering state given by Eq.~\eqref{tot_scatt_alpha:eq} for the Aharonov--Bohm scattering problem in isotropic multiband systems. We provided rigorous proofs that this AB scattering state indeed satisfies the Schr\"odinger equation.  We also showed that for a system of spinless free particles our AB scattering state reduces to the form obtained first by Aharonov and Bohm. Moreover, we found a remarkable result, namely the differential scattering cross section is the same for all isotropic multiband systems. As an application, for a few specific isotropic multiband systems, we carried out the complex integrals given in Eq.~\eqref{tot_scatt_alpha:eq}. In the Supplemental Material, we visualized the wave functions for the systems listed in Table~\ref{systems:table} in a similar way as in Ref.~\cite{Berry_streamline_501_2017}. We believe that our work provides a better insight into the famous Aharonov--Bohm effect for multiband systems, and could be experimentally applicable, for example, to strained or gapped graphenes, or to tomographic
imaging~\cite{Valagiannopoulos_tomography:cikk}. 
Finally, the extension of our integral representation of the AB scattering state to anisotropic multiband systems, or to the case of a finite radius solenoid is a further relevant research direction.

We thank M.~Berry, C.~Lambert, B.~D\'ora, and Gy.~D\'avid  for valuable discussions.  
This research was supported by the Ministry of Culture and Innovation and the National Research,
Development and Innovation Office within the Quantum Information National Laboratory of Hungary
(Grant No. 2022-2.1.1-NL-2022-00004), by the ÚNKP-22-3 New National Excellence Program of the Ministry for Culture and Innovation from the Source of the National Research, Development and Innovation Fund, and by the Innovation Office (NKFIH) through Grant Nos. K134437.	

%\bibliography{cikkek_AB.bib}

%\bibliographystyle{/home/cserti/Dropbox/programok/tex/prsty}
%\bibliography{/home/cserti/Dropbox/AD/BibTeX/cikkek.bib}

\begin{thebibliography}{48}
	\expandafter\ifx\csname natexlab\endcsname\relax\def\natexlab#1{#1}\fi
	\expandafter\ifx\csname bibnamefont\endcsname\relax
	\def\bibnamefont#1{#1}\fi
	\expandafter\ifx\csname bibfnamefont\endcsname\relax
	\def\bibfnamefont#1{#1}\fi
	\expandafter\ifx\csname citenamefont\endcsname\relax
	\def\citenamefont#1{#1}\fi
	\expandafter\ifx\csname url\endcsname\relax
	\def\url#1{\texttt{#1}}\fi
	\expandafter\ifx\csname urlprefix\endcsname\relax\def\urlprefix{URL }\fi
	\providecommand{\bibinfo}[2]{#2}
	\providecommand{\eprint}[2][]{\url{#2}}
	
	\bibitem[{\citenamefont{Aharonov and Bohm}(1959)}]{Aharonov-Bohm:cikk}
	\bibinfo{author}{\bibfnamefont{Y.}~\bibnamefont{Aharonov}} \bibnamefont{and}
	\bibinfo{author}{\bibfnamefont{D.}~\bibnamefont{Bohm}},
	\bibinfo{journal}{Phys. Rev.} \textbf{\bibinfo{volume}{115}},
	\bibinfo{pages}{485} (\bibinfo{year}{1959}).
	
	\bibitem[{\citenamefont{Chambers}(1960)}]{PhysRevLett.5.3}
	\bibinfo{author}{\bibfnamefont{R.~G.} \bibnamefont{Chambers}},
	\bibinfo{journal}{Phys. Rev. Lett.} \textbf{\bibinfo{volume}{5}},
	\bibinfo{pages}{3} (\bibinfo{year}{1960}).
	
	\bibitem[{\citenamefont{Tonomura et~al.}(1982)\citenamefont{Tonomura, Matsuda,
			Suzuki, Fukuhara, Osakabe, Umezaki, Endo, Shinagawa, Sugita, and
			Fujiwara}}]{Tonomura_PhysRevLett.48.1443}
	\bibinfo{author}{\bibfnamefont{A.}~\bibnamefont{Tonomura}},
	\bibinfo{author}{\bibfnamefont{T.}~\bibnamefont{Matsuda}},
	\bibinfo{author}{\bibfnamefont{R.}~\bibnamefont{Suzuki}},
	\bibinfo{author}{\bibfnamefont{A.}~\bibnamefont{Fukuhara}},
	\bibinfo{author}{\bibfnamefont{N.}~\bibnamefont{Osakabe}},
	\bibinfo{author}{\bibfnamefont{H.}~\bibnamefont{Umezaki}},
	\bibinfo{author}{\bibfnamefont{J.}~\bibnamefont{Endo}},
	\bibinfo{author}{\bibfnamefont{K.}~\bibnamefont{Shinagawa}},
	\bibinfo{author}{\bibfnamefont{Y.}~\bibnamefont{Sugita}}, \bibnamefont{and}
	\bibinfo{author}{\bibfnamefont{H.}~\bibnamefont{Fujiwara}},
	\bibinfo{journal}{Phys. Rev. Lett.} \textbf{\bibinfo{volume}{48}},
	\bibinfo{pages}{1443} (\bibinfo{year}{1982}).
	
	\bibitem[{\citenamefont{Webb et~al.}(1985)\citenamefont{Webb, Washburn, Umbach,
			and Laibowitz}}]{Webb_PhysRevLett.54.2696}
	\bibinfo{author}{\bibfnamefont{R.~A.} \bibnamefont{Webb}},
	\bibinfo{author}{\bibfnamefont{S.}~\bibnamefont{Washburn}},
	\bibinfo{author}{\bibfnamefont{C.~P.} \bibnamefont{Umbach}},
	\bibnamefont{and} \bibinfo{author}{\bibfnamefont{R.~B.}
		\bibnamefont{Laibowitz}}, \bibinfo{journal}{Phys. Rev. Lett.}
	\textbf{\bibinfo{volume}{54}}, \bibinfo{pages}{2696} (\bibinfo{year}{1985}).
	
	\bibitem[{\citenamefont{Fuhrer et~al.}(2001)\citenamefont{Fuhrer, L{\"u}scher,
			Ihn, Heinzel, Ensslin, Wegscheider, and Bichler}}]{Fuhrer_Ensslin_2001}
	\bibinfo{author}{\bibfnamefont{A.}~\bibnamefont{Fuhrer}},
	\bibinfo{author}{\bibfnamefont{S.}~\bibnamefont{L{\"u}scher}},
	\bibinfo{author}{\bibfnamefont{T.}~\bibnamefont{Ihn}},
	\bibinfo{author}{\bibfnamefont{T.}~\bibnamefont{Heinzel}},
	\bibinfo{author}{\bibfnamefont{K.}~\bibnamefont{Ensslin}},
	\bibinfo{author}{\bibfnamefont{W.}~\bibnamefont{Wegscheider}},
	\bibnamefont{and} \bibinfo{author}{\bibfnamefont{M.}~\bibnamefont{Bichler}},
	\bibinfo{journal}{Nature} \textbf{\bibinfo{volume}{413}},
	\bibinfo{pages}{822} (\bibinfo{year}{2001}).
	
	\bibitem[{\citenamefont{Sigrist et~al.}(2004)\citenamefont{Sigrist, Fuhrer,
			Ihn, Ensslin, Ulloa, Wegscheider, and
			Bichler}}]{Ensslin_PhysRevLett.93.066802}
	\bibinfo{author}{\bibfnamefont{M.}~\bibnamefont{Sigrist}},
	\bibinfo{author}{\bibfnamefont{A.}~\bibnamefont{Fuhrer}},
	\bibinfo{author}{\bibfnamefont{T.}~\bibnamefont{Ihn}},
	\bibinfo{author}{\bibfnamefont{K.}~\bibnamefont{Ensslin}},
	\bibinfo{author}{\bibfnamefont{S.~E.} \bibnamefont{Ulloa}},
	\bibinfo{author}{\bibfnamefont{W.}~\bibnamefont{Wegscheider}},
	\bibnamefont{and} \bibinfo{author}{\bibfnamefont{M.}~\bibnamefont{Bichler}},
	\bibinfo{journal}{Phys. Rev. Lett.} \textbf{\bibinfo{volume}{93}},
	\bibinfo{pages}{066802} (\bibinfo{year}{2004}).
	
	\bibitem[{\citenamefont{Ensslin}(2005)}]{ENSSLIN2005585}
	\bibinfo{author}{\bibfnamefont{K.}~\bibnamefont{Ensslin}}, in
	\emph{\bibinfo{booktitle}{Nanophysics: Coherence and Transport}}, edited by
	\bibinfo{editor}{\bibfnamefont{H.}~\bibnamefont{Bouchiat}},
	\bibinfo{editor}{\bibfnamefont{Y.}~\bibnamefont{Gefen}},
	\bibinfo{editor}{\bibfnamefont{S.}~\bibnamefont{Guéron}},
	\bibinfo{editor}{\bibfnamefont{G.}~\bibnamefont{Montambaux}},
	\bibnamefont{and} \bibinfo{editor}{\bibfnamefont{J.}~\bibnamefont{Dalibard}}
	(\bibinfo{publisher}{Elsevier}, \bibinfo{year}{2005}),
	vol.~\bibinfo{volume}{81} of \emph{\bibinfo{series}{Les Houches}}, pp.
	\bibinfo{pages}{585--586}.
	
	\bibitem[{\citenamefont{Olariu and Popescu}(1985)}]{Popescu_RevModPhys.57.339}
	\bibinfo{author}{\bibfnamefont{S.}~\bibnamefont{Olariu}} \bibnamefont{and}
	\bibinfo{author}{\bibfnamefont{I.~I.} \bibnamefont{Popescu}},
	\bibinfo{journal}{Rev. Mod. Phys.} \textbf{\bibinfo{volume}{57}},
	\bibinfo{pages}{339} (\bibinfo{year}{1985}).
	
	\bibitem[{\citenamefont{Hagen}(1990{\natexlab{a}})}]{Hagen_PhysRevD.41.2015}
	\bibinfo{author}{\bibfnamefont{C.~R.} \bibnamefont{Hagen}},
	\bibinfo{journal}{Phys. Rev. D} \textbf{\bibinfo{volume}{41}},
	\bibinfo{pages}{2015} (\bibinfo{year}{1990}{\natexlab{a}}).
	
	\bibitem[{\citenamefont{Berry}(1980)}]{Berry_1980}
	\bibinfo{author}{\bibfnamefont{M.~V.} \bibnamefont{Berry}},
	\bibinfo{journal}{European Journal of Physics} \textbf{\bibinfo{volume}{1}},
	\bibinfo{pages}{240} (\bibinfo{year}{1980}).
	
	\bibitem[{\citenamefont{Alford and
			Wilczek}(1989)}]{Wilczek_PhysRevLett.62.1071}
	\bibinfo{author}{\bibfnamefont{M.~G.} \bibnamefont{Alford}} \bibnamefont{and}
	\bibinfo{author}{\bibfnamefont{F.}~\bibnamefont{Wilczek}},
	\bibinfo{journal}{Phys. Rev. Lett.} \textbf{\bibinfo{volume}{62}},
	\bibinfo{pages}{1071} (\bibinfo{year}{1989}).
	
	\bibitem[{\citenamefont{Gerbert}(1989)}]{Gerbert_PhysRevD.40.1346}
	\bibinfo{author}{\bibfnamefont{P.~d.~S.} \bibnamefont{Gerbert}},
	\bibinfo{journal}{Phys. Rev. D} \textbf{\bibinfo{volume}{40}},
	\bibinfo{pages}{1346} (\bibinfo{year}{1989}).
	
	\bibitem[{\citenamefont{Hagen}(1990{\natexlab{b}})}]{Hagen_PhysRevLett.64.503}
	\bibinfo{author}{\bibfnamefont{C.~R.} \bibnamefont{Hagen}},
	\textbf{\bibinfo{volume}{64}}, \bibinfo{pages}{503}
	(\bibinfo{year}{1990}{\natexlab{b}}).
	
	\bibitem[{\citenamefont{Hagen and Ramaswamy}(1990)}]{Hagen_PhysRevD.42.3524}
	\bibinfo{author}{\bibfnamefont{C.~R.} \bibnamefont{Hagen}} \bibnamefont{and}
	\bibinfo{author}{\bibfnamefont{S.}~\bibnamefont{Ramaswamy}},
	\bibinfo{journal}{Phys. Rev. D} \textbf{\bibinfo{volume}{42}},
	\bibinfo{pages}{3524} (\bibinfo{year}{1990}).
	
	\bibitem[{\citenamefont{Slobodeniuk et~al.}(2010)\citenamefont{Slobodeniuk,
			Sharapov, and Loktev}}]{PhysRevB.82.075316}
	\bibinfo{author}{\bibfnamefont{A.~O.} \bibnamefont{Slobodeniuk}},
	\bibinfo{author}{\bibfnamefont{S.~G.} \bibnamefont{Sharapov}},
	\bibnamefont{and} \bibinfo{author}{\bibfnamefont{V.~M.}
		\bibnamefont{Loktev}}, \bibinfo{journal}{Phys. Rev. B}
	\textbf{\bibinfo{volume}{82}}, \bibinfo{pages}{075316}
	(\bibinfo{year}{2010}).
	
	\bibitem[{\citenamefont{Slobodeniuk et~al.}(2011)\citenamefont{Slobodeniuk,
			Sharapov, and Loktev}}]{PhysRevB.84.125306}
	\bibinfo{author}{\bibfnamefont{A.~O.} \bibnamefont{Slobodeniuk}},
	\bibinfo{author}{\bibfnamefont{S.~G.} \bibnamefont{Sharapov}},
	\bibnamefont{and} \bibinfo{author}{\bibfnamefont{V.~M.}
		\bibnamefont{Loktev}}, \bibinfo{journal}{Phys. Rev. B}
	\textbf{\bibinfo{volume}{84}}, \bibinfo{pages}{125306}
	(\bibinfo{year}{2011}).
	
	\bibitem[{\citenamefont{Girotti and Romero}(1997)}]{Girotti_1997}
	\bibinfo{author}{\bibfnamefont{H.~O.} \bibnamefont{Girotti}} \bibnamefont{and}
	\bibinfo{author}{\bibfnamefont{F.~F.} \bibnamefont{Romero}},
	\bibinfo{journal}{Europhysics Letters ({EPL})} \textbf{\bibinfo{volume}{37}},
	\bibinfo{pages}{165} (\bibinfo{year}{1997}).
	
	\bibitem[{\citenamefont{Recher et~al.}(2007)\citenamefont{Recher, Trauzettel,
			Rycerz, Blanter, Beenakker, and
			Morpurgo}}]{Carlo_AB_Graphene_PhysRevB.76.235404}
	\bibinfo{author}{\bibfnamefont{P.}~\bibnamefont{Recher}},
	\bibinfo{author}{\bibfnamefont{B.}~\bibnamefont{Trauzettel}},
	\bibinfo{author}{\bibfnamefont{A.}~\bibnamefont{Rycerz}},
	\bibinfo{author}{\bibfnamefont{Y.~M.} \bibnamefont{Blanter}},
	\bibinfo{author}{\bibfnamefont{C.~W.~J.} \bibnamefont{Beenakker}},
	\bibnamefont{and} \bibinfo{author}{\bibfnamefont{A.~F.}
		\bibnamefont{Morpurgo}}, \bibinfo{journal}{Phys. Rev. B}
	\textbf{\bibinfo{volume}{76}}, \bibinfo{pages}{235404}
	(\bibinfo{year}{2007}).
	
	\bibitem[{\citenamefont{Russo et~al.}(2008)\citenamefont{Russo, Oostinga,
			Wehenkel, Heersche, Sobhani, Vandersypen, and
			Morpurgo}}]{AB_Graphene_PhysRevB.77.085413}
	\bibinfo{author}{\bibfnamefont{S.}~\bibnamefont{Russo}},
	\bibinfo{author}{\bibfnamefont{J.~B.} \bibnamefont{Oostinga}},
	\bibinfo{author}{\bibfnamefont{D.}~\bibnamefont{Wehenkel}},
	\bibinfo{author}{\bibfnamefont{H.~B.} \bibnamefont{Heersche}},
	\bibinfo{author}{\bibfnamefont{S.~S.} \bibnamefont{Sobhani}},
	\bibinfo{author}{\bibfnamefont{L.~M.~K.} \bibnamefont{Vandersypen}},
	\bibnamefont{and} \bibinfo{author}{\bibfnamefont{A.~F.}
		\bibnamefont{Morpurgo}}, \bibinfo{journal}{Phys. Rev. B}
	\textbf{\bibinfo{volume}{77}}, \bibinfo{pages}{085413}
	(\bibinfo{year}{2008}).
	
	\bibitem[{\citenamefont{Huefner et~al.}(2009)\citenamefont{Huefner, Molitor,
			Jacobsen, Pioda, Stampfer, Ensslin, and Ihn}}]{AB_Graphene_Huefner_2009}
	\bibinfo{author}{\bibfnamefont{M.}~\bibnamefont{Huefner}},
	\bibinfo{author}{\bibfnamefont{F.}~\bibnamefont{Molitor}},
	\bibinfo{author}{\bibfnamefont{A.}~\bibnamefont{Jacobsen}},
	\bibinfo{author}{\bibfnamefont{A.}~\bibnamefont{Pioda}},
	\bibinfo{author}{\bibfnamefont{C.}~\bibnamefont{Stampfer}},
	\bibinfo{author}{\bibfnamefont{K.}~\bibnamefont{Ensslin}}, \bibnamefont{and}
	\bibinfo{author}{\bibfnamefont{T.}~\bibnamefont{Ihn}},
	\bibinfo{journal}{Phys. Status Solidi B} \textbf{\bibinfo{volume}{246}},
	\bibinfo{pages}{2756} (\bibinfo{year}{2009}).
	
	\bibitem[{\citenamefont{Huefner et~al.}(2010)\citenamefont{Huefner, Molitor,
			Jacobsen, Pioda, Stampfer, Ensslin, and Ihn}}]{AB_Graphene_Huefner_2010}
	\bibinfo{author}{\bibfnamefont{M.}~\bibnamefont{Huefner}},
	\bibinfo{author}{\bibfnamefont{F.}~\bibnamefont{Molitor}},
	\bibinfo{author}{\bibfnamefont{A.}~\bibnamefont{Jacobsen}},
	\bibinfo{author}{\bibfnamefont{A.}~\bibnamefont{Pioda}},
	\bibinfo{author}{\bibfnamefont{C.}~\bibnamefont{Stampfer}},
	\bibinfo{author}{\bibfnamefont{K.}~\bibnamefont{Ensslin}}, \bibnamefont{and}
	\bibinfo{author}{\bibfnamefont{T.}~\bibnamefont{Ihn}}, \bibinfo{journal}{New
		Journal of Physics} \textbf{\bibinfo{volume}{12}}, \bibinfo{pages}{043054}
	(\bibinfo{year}{2010}).
	
	\bibitem[{\citenamefont{Wurm et~al.}(2010)\citenamefont{Wurm, Wimmer, Baranger,
			and Richter}}]{AB_graphene_Wurm_Richter_2010}
	\bibinfo{author}{\bibfnamefont{J.}~\bibnamefont{Wurm}},
	\bibinfo{author}{\bibfnamefont{M.}~\bibnamefont{Wimmer}},
	\bibinfo{author}{\bibfnamefont{H.~U.} \bibnamefont{Baranger}},
	\bibnamefont{and} \bibinfo{author}{\bibfnamefont{K.}~\bibnamefont{Richter}},
	\bibinfo{journal}{Semiconductor Science and Technology}
	\textbf{\bibinfo{volume}{25}}, \bibinfo{pages}{034003}
	(\bibinfo{year}{2010}).
	
	\bibitem[{\citenamefont{de~Juan et~al.}(2011)\citenamefont{de~Juan, Cortijo,
			Vozmediano, and Cano}}]{AB_graphene_Vozmediano_deJuan_2011}
	\bibinfo{author}{\bibfnamefont{F.}~\bibnamefont{de~Juan}},
	\bibinfo{author}{\bibfnamefont{A.}~\bibnamefont{Cortijo}},
	\bibinfo{author}{\bibfnamefont{M.~A.~H.} \bibnamefont{Vozmediano}},
	\bibnamefont{and} \bibinfo{author}{\bibfnamefont{A.}~\bibnamefont{Cano}},
	\bibinfo{journal}{Nature Physics} \textbf{\bibinfo{volume}{7}},
	\bibinfo{pages}{810} (\bibinfo{year}{2011}).
	
	\bibitem[{\citenamefont{Park}(2017)}]{PARK20171831}
	\bibinfo{author}{\bibfnamefont{C.-S.} \bibnamefont{Park}},
	\bibinfo{journal}{Physics Letters A} \textbf{\bibinfo{volume}{381}},
	\bibinfo{pages}{1831} (\bibinfo{year}{2017}).
	
	\bibitem[{\citenamefont{Ramezani~Masir and
			Peeters}(2013)}]{Ramezani_Peeters_scatt_2013}
	\bibinfo{author}{\bibfnamefont{M.}~\bibnamefont{Ramezani~Masir}}
	\bibnamefont{and} \bibinfo{author}{\bibfnamefont{F.}~\bibnamefont{Peeters}},
	\bibinfo{journal}{Journal of Computational Electronics}
	\textbf{\bibinfo{volume}{12}}, \bibinfo{pages}{115} (\bibinfo{year}{2013}).
	
	\bibitem[{\citenamefont{Rycerz and Suszalski}(2020)}]{Rycerz_PRB.101.245429}
	\bibinfo{author}{\bibfnamefont{A.}~\bibnamefont{Rycerz}} \bibnamefont{and}
	\bibinfo{author}{\bibfnamefont{D.}~\bibnamefont{Suszalski}},
	\bibinfo{journal}{Phys. Rev. B} \textbf{\bibinfo{volume}{101}},
	\bibinfo{pages}{245429} (\bibinfo{year}{2020}).
	
	\bibitem[{\citenamefont{Zazunov et~al.}(2010)\citenamefont{Zazunov, Kundu,
			H\"utten, and
			Egger}}]{Dirac_in_top_insul_and_graphene_Egger_PhysRevB.82.155431}
	\bibinfo{author}{\bibfnamefont{A.}~\bibnamefont{Zazunov}},
	\bibinfo{author}{\bibfnamefont{A.}~\bibnamefont{Kundu}},
	\bibinfo{author}{\bibfnamefont{A.}~\bibnamefont{H\"utten}}, \bibnamefont{and}
	\bibinfo{author}{\bibfnamefont{R.}~\bibnamefont{Egger}},
	\bibinfo{journal}{Phys. Rev. B} \textbf{\bibinfo{volume}{82}},
	\bibinfo{pages}{155431} (\bibinfo{year}{2010}).
	
	\bibitem[{\citenamefont{Gerbert and
			Jackiw}(1989)}]{Gerbert_Jackiw_MIT_Report_1989}
	\bibinfo{author}{\bibfnamefont{P.~d.~S.} \bibnamefont{Gerbert}}
	\bibnamefont{and} \bibinfo{author}{\bibfnamefont{R.}~\bibnamefont{Jackiw}},
	\bibinfo{journal}{Commun. Math. Phys.} \textbf{\bibinfo{volume}{124}},
	\bibinfo{pages}{229} (\bibinfo{year}{1989}).
	
	\bibitem[{\citenamefont{Gradshteyn and Ryzhik}(1994)}]{Grads:book}
	\bibinfo{author}{\bibfnamefont{I.~S.} \bibnamefont{Gradshteyn}}
	\bibnamefont{and} \bibinfo{author}{\bibfnamefont{I.~M.}
		\bibnamefont{Ryzhik}}, \emph{\bibinfo{title}{Table of Integrals, Series, and
			Products}} (\bibinfo{publisher}{Academic Press}, \bibinfo{address}{London,
		UK}, \bibinfo{year}{1994}), \bibinfo{edition}{5th} ed.
	
	\bibitem[{\citenamefont{Berry et~al.}(1980)\citenamefont{Berry, Chambers,
			Large, Upstill, and Walmsley}}]{Berry_Chambers_1980}
	\bibinfo{author}{\bibfnamefont{M.~V.} \bibnamefont{Berry}},
	\bibinfo{author}{\bibfnamefont{R.~G.} \bibnamefont{Chambers}},
	\bibinfo{author}{\bibfnamefont{M.~D.} \bibnamefont{Large}},
	\bibinfo{author}{\bibfnamefont{C.}~\bibnamefont{Upstill}}, \bibnamefont{and}
	\bibinfo{author}{\bibfnamefont{J.~C.} \bibnamefont{Walmsley}},
	\bibinfo{journal}{European Journal of Physics} \textbf{\bibinfo{volume}{1}},
	\bibinfo{pages}{154} (\bibinfo{year}{1980}).
	
	\bibitem[{\citenamefont{Arfken and Weber}(1995)}]{Arfken}
	\bibinfo{author}{\bibfnamefont{G.~B.} \bibnamefont{Arfken}} \bibnamefont{and}
	\bibinfo{author}{\bibfnamefont{H.~J.} \bibnamefont{Weber}},
	\emph{\bibinfo{title}{Mathematical Methods for Physicists}}
	(\bibinfo{publisher}{Academic Press}, \bibinfo{address}{San Diego, CA},
	\bibinfo{year}{1995}), \bibinfo{edition}{4th} ed.
	
	\bibitem[{\citenamefont{Datta}(1995)}]{datta_1995}
	\bibinfo{author}{\bibfnamefont{S.}~\bibnamefont{Datta}},
	\emph{\bibinfo{title}{Electronic Transport in Mesoscopic Systems}}, Cambridge
	Studies in Semiconductor Physics and Microelectronic Engineering
	(\bibinfo{publisher}{Cambridge University Press}, \bibinfo{year}{1995}).
	
	\bibitem[{\citenamefont{Novoselov et~al.}(2005)\citenamefont{Novoselov, Geim,
			Morozov, Jiang, Katsnelson, Grigorieva, Dubonos, and Firsov}}]{Novoselov2005}
	\bibinfo{author}{\bibfnamefont{K.~S.} \bibnamefont{Novoselov}},
	\bibinfo{author}{\bibfnamefont{A.~K.} \bibnamefont{Geim}},
	\bibinfo{author}{\bibfnamefont{S.~V.} \bibnamefont{Morozov}},
	\bibinfo{author}{\bibfnamefont{D.}~\bibnamefont{Jiang}},
	\bibinfo{author}{\bibfnamefont{M.~I.} \bibnamefont{Katsnelson}},
	\bibinfo{author}{\bibfnamefont{I.~V.} \bibnamefont{Grigorieva}},
	\bibinfo{author}{\bibfnamefont{S.~V.} \bibnamefont{Dubonos}},
	\bibnamefont{and} \bibinfo{author}{\bibfnamefont{A.~A.}
		\bibnamefont{Firsov}}, \bibinfo{journal}{Nature}
	\textbf{\bibinfo{volume}{438}}, \bibinfo{pages}{197} (\bibinfo{year}{2005}).
	
	\bibitem[{\citenamefont{Zhang et~al.}(2005)\citenamefont{Zhang, Tan, Stormer,
			and Kim}}]{Zhang2005}
	\bibinfo{author}{\bibfnamefont{Y.}~\bibnamefont{Zhang}},
	\bibinfo{author}{\bibfnamefont{Y.-W.} \bibnamefont{Tan}},
	\bibinfo{author}{\bibfnamefont{H.~L.} \bibnamefont{Stormer}},
	\bibnamefont{and} \bibinfo{author}{\bibfnamefont{P.}~\bibnamefont{Kim}},
	\bibinfo{journal}{Nature} \textbf{\bibinfo{volume}{438}},
	\bibinfo{pages}{201} (\bibinfo{year}{2005}).
	
	\bibitem[{\citenamefont{DiVincenzo and
			Mele}(1984)}]{DiVincenzo_Mele_PRB_29_1685:ref}
	\bibinfo{author}{\bibfnamefont{D.~P.} \bibnamefont{DiVincenzo}}
	\bibnamefont{and} \bibinfo{author}{\bibfnamefont{E.~J.} \bibnamefont{Mele}},
	\bibinfo{journal}{Phys. Rev. B} \textbf{\bibinfo{volume}{29}},
	\bibinfo{pages}{1685} (\bibinfo{year}{1984}).
	
	\bibitem[{\citenamefont{Castro~Neto et~al.}(2009)\citenamefont{Castro~Neto,
			Guinea, Peres, Novoselov, and Geim}}]{RevModPhys.81.109}
	\bibinfo{author}{\bibfnamefont{A.~H.} \bibnamefont{Castro~Neto}},
	\bibinfo{author}{\bibfnamefont{F.}~\bibnamefont{Guinea}},
	\bibinfo{author}{\bibfnamefont{N.~M.~R.} \bibnamefont{Peres}},
	\bibinfo{author}{\bibfnamefont{K.~S.} \bibnamefont{Novoselov}},
	\bibnamefont{and} \bibinfo{author}{\bibfnamefont{A.~K.} \bibnamefont{Geim}},
	\bibinfo{journal}{Rev. Mod. Phys.} \textbf{\bibinfo{volume}{81}},
	\bibinfo{pages}{109} (\bibinfo{year}{2009}).
	
	\bibitem[{\citenamefont{Novoselov et~al.}({2006})\citenamefont{Novoselov,
			McCann, Morozov, Fal'ko, Katsnelson, Zeitler, Jiang, Schedin, and
			Geim}}]{Novoselov_Hall:ref}
	\bibinfo{author}{\bibfnamefont{K.}~\bibnamefont{Novoselov}},
	\bibinfo{author}{\bibfnamefont{E.}~\bibnamefont{McCann}},
	\bibinfo{author}{\bibfnamefont{S.}~\bibnamefont{Morozov}},
	\bibinfo{author}{\bibfnamefont{V.}~\bibnamefont{Fal'ko}},
	\bibinfo{author}{\bibfnamefont{M.}~\bibnamefont{Katsnelson}},
	\bibinfo{author}{\bibfnamefont{U.}~\bibnamefont{Zeitler}},
	\bibinfo{author}{\bibfnamefont{D.}~\bibnamefont{Jiang}},
	\bibinfo{author}{\bibfnamefont{F.}~\bibnamefont{Schedin}}, \bibnamefont{and}
	\bibinfo{author}{\bibfnamefont{A.}~\bibnamefont{Geim}},
	\bibinfo{journal}{Nature Phys.} \textbf{\bibinfo{volume}{{2}}},
	\bibinfo{pages}{177} (\bibinfo{year}{{2006}}).
	
	\bibitem[{\citenamefont{McCann and Fal'ko}(2006)}]{mccann:086805}
	\bibinfo{author}{\bibfnamefont{E.}~\bibnamefont{McCann}} \bibnamefont{and}
	\bibinfo{author}{\bibfnamefont{V.~I.} \bibnamefont{Fal'ko}},
	\bibinfo{journal}{Phys. Rev. Lett.} \textbf{\bibinfo{volume}{96}},
	\bibinfo{eid}{086805} (\bibinfo{year}{2006}).
	
	\bibitem[{\citenamefont{Rashba}(1960)}]{Rashba_cikk_1960}
	\bibinfo{author}{\bibfnamefont{E.}~\bibnamefont{Rashba}},
	\bibinfo{journal}{Sov. Phys.-Solid State} \textbf{\bibinfo{volume}{2}},
	\bibinfo{pages}{1109} (\bibinfo{year}{1960}).
	
	\bibitem[{\citenamefont{Bychkov and Rashba}(1984)}]{Yu_A_Bychkov_1984}
	\bibinfo{author}{\bibfnamefont{Y.~A.} \bibnamefont{Bychkov}} \bibnamefont{and}
	\bibinfo{author}{\bibfnamefont{E.~I.} \bibnamefont{Rashba}},
	\bibinfo{journal}{Journal of Physics C: Solid State Physics}
	\textbf{\bibinfo{volume}{17}}, \bibinfo{pages}{6039} (\bibinfo{year}{1984}).
	
	\bibitem[{\citenamefont{\ifmmode \check{Z}\else
			\v{Z}\fi{}uti\ifmmode~\acute{c}\else \'{c}\fi{}
			et~al.}(2004)\citenamefont{\ifmmode \check{Z}\else
			\v{Z}\fi{}uti\ifmmode~\acute{c}\else \'{c}\fi{}, Fabian, and
			Das~Sarma}}]{Fabian_RevModPhys.76.323}
	\bibinfo{author}{\bibfnamefont{I.}~\bibnamefont{\ifmmode \check{Z}\else
			\v{Z}\fi{}uti\ifmmode~\acute{c}\else \'{c}\fi{}}},
	\bibinfo{author}{\bibfnamefont{J.}~\bibnamefont{Fabian}}, \bibnamefont{and}
	\bibinfo{author}{\bibfnamefont{S.}~\bibnamefont{Das~Sarma}},
	\bibinfo{journal}{Rev. Mod. Phys.} \textbf{\bibinfo{volume}{76}},
	\bibinfo{pages}{323} (\bibinfo{year}{2004}).
	
	\bibitem[{\citenamefont{Winkler}(2003)}]{Winkler:684956}
	\bibinfo{author}{\bibfnamefont{R.}~\bibnamefont{Winkler}},
	\emph{\bibinfo{title}{{Spin-orbit coupling effects in two-dimensional
				electron and hole systems}}}, Springer Tracts in Modern Physics
	(\bibinfo{publisher}{Springer}, \bibinfo{address}{Berlin},
	\bibinfo{year}{2003}).
	
	\bibitem[{\citenamefont{Urban et~al.}(2011)\citenamefont{Urban, Bercioux,
			Wimmer, and H\"ausler}}]{PhysRevB.84.115136}
	\bibinfo{author}{\bibfnamefont{D.~F.} \bibnamefont{Urban}},
	\bibinfo{author}{\bibfnamefont{D.}~\bibnamefont{Bercioux}},
	\bibinfo{author}{\bibfnamefont{M.}~\bibnamefont{Wimmer}}, \bibnamefont{and}
	\bibinfo{author}{\bibfnamefont{W.}~\bibnamefont{H\"ausler}},
	\bibinfo{journal}{Phys. Rev. B} \textbf{\bibinfo{volume}{84}},
	\bibinfo{pages}{115136} (\bibinfo{year}{2011}).
	
	\bibitem[{\citenamefont{Shen et~al.}(2010)\citenamefont{Shen, Shao, Wang, and
			Xing}}]{PhysRevB.81.041410}
	\bibinfo{author}{\bibfnamefont{R.}~\bibnamefont{Shen}},
	\bibinfo{author}{\bibfnamefont{L.~B.} \bibnamefont{Shao}},
	\bibinfo{author}{\bibfnamefont{B.}~\bibnamefont{Wang}}, \bibnamefont{and}
	\bibinfo{author}{\bibfnamefont{D.~Y.} \bibnamefont{Xing}},
	\bibinfo{journal}{Phys. Rev. B} \textbf{\bibinfo{volume}{81}},
	\bibinfo{pages}{041410} (\bibinfo{year}{2010}).
	
	\bibitem[{\citenamefont{Bercioux et~al.}(2009)\citenamefont{Bercioux, Urban,
			Grabert, and H\"ausler}}]{PhysRevA.80.063603}
	\bibinfo{author}{\bibfnamefont{D.}~\bibnamefont{Bercioux}},
	\bibinfo{author}{\bibfnamefont{D.~F.} \bibnamefont{Urban}},
	\bibinfo{author}{\bibfnamefont{H.}~\bibnamefont{Grabert}}, \bibnamefont{and}
	\bibinfo{author}{\bibfnamefont{W.}~\bibnamefont{H\"ausler}},
	\bibinfo{journal}{Phys. Rev. A} \textbf{\bibinfo{volume}{80}},
	\bibinfo{pages}{063603} (\bibinfo{year}{2009}).
	
	\bibitem[{\citenamefont{D\'ora et~al.}(2011)\citenamefont{D\'ora, Kailasvuori,
			and Moessner}}]{PhysRevB.84.195422}
	\bibinfo{author}{\bibfnamefont{B.}~\bibnamefont{D\'ora}},
	\bibinfo{author}{\bibfnamefont{J.}~\bibnamefont{Kailasvuori}},
	\bibnamefont{and} \bibinfo{author}{\bibfnamefont{R.}~\bibnamefont{Moessner}},
	\bibinfo{journal}{Phys. Rev. B} \textbf{\bibinfo{volume}{84}},
	\bibinfo{pages}{195422} (\bibinfo{year}{2011}).
	
	\bibitem[{\citenamefont{Berry}(2017)}]{Berry_streamline_501_2017}
	\bibinfo{author}{\bibfnamefont{M.~V.} \bibnamefont{Berry}},
	\bibinfo{journal}{Journal of Physics A: Mathematical and Theoretical}
	\textbf{\bibinfo{volume}{50}}, \bibinfo{pages}{43LT01}
	(\bibinfo{year}{2017}).
	
	\bibitem[{\citenamefont{Valagiannopoulos
			et~al.}(2018)\citenamefont{Valagiannopoulos, Marengo, Dimakis, and
			Alù}}]{Valagiannopoulos_tomography:cikk}
	\bibinfo{author}{\bibfnamefont{C.}~\bibnamefont{Valagiannopoulos}},
	\bibinfo{author}{\bibfnamefont{E.~A.} \bibnamefont{Marengo}},
	\bibinfo{author}{\bibfnamefont{A.~G.} \bibnamefont{Dimakis}},
	\bibnamefont{and} \bibinfo{author}{\bibfnamefont{A.}~\bibnamefont{Alù}},
	\bibinfo{journal}{IET Microwaves, Antennas \& Propagation}
	\textbf{\bibinfo{volume}{12}}, \bibinfo{pages}{1890} (\bibinfo{year}{2018}).
	
\end{thebibliography}

%\begin{thebibliography}{}
%\end{thebibliography}

\end{document}